\documentclass{PoS}

\newcommand{\be}{\begin{equation}}
\newcommand{\ee}{\end{equation}}
\newcommand{\bary}{\begin{eqnarray}}
\newcommand{\eary}{\end{eqnarray}}

\newcommand{\msun}{\mbox{$M_\odot$}}

\def\be{\begin{eqnarray}}
\def\ee{\end{eqnarray}}
\def\bi{\begin{itemize}}
\def\ei{\end{itemize}}
\def\lsim{\mathrel{\rlap{\lower3pt\hbox{\hskip1pt$\sim$}}
     \raise1pt\hbox{$<$}}} %less than or approx. symbol
\def\gsim{\mathrel{\rlap{\lower3pt\hbox{\hskip1pt$\sim$}}
     \raise1pt\hbox{$>$}}} %greater than or approx. symbol

\title{Reviewing the case of the atypical central-engine activity in GRB 110709B}

\ShortTitle{GRB110709B}

\author{\speaker{Nissim Fraija}\thanks{Luc Binette postdoctoral scholarship.}\\
        Instituto de Astronom\'ia - UNAM\\
        E-mail: \email{nifraija@astro.unam.mx}}

\author{Enrique Moreno-M\'endez\\
        Instituto de Astronom\'ia - UNAM\\
        Universidad An\'ahuac M\'exico Sur\\
        E-mail: \email{enriquemm@astro.unam.mx}}

\author{Barbara Patricelli\\
	Universit\`a di Pisa, I-56127 Pisa, Italy \\
	INFN, Sezione di Pisa, I-56127 Pisa, Italy\\
        Instituto de Astronom\'ia - UNAM\\
        E-mail: \email{barbara.patricelli@pi.infn.it}}

\abstract{The unusual GRB 110709B triggered Swift/BAT twice, with a time difference of $\sim 11$ minutes. 
Its light curve presented three noticeable peaks but only two were originally identified. 
In this work, we describe each peak as due to a different central-engine phase:  the first one is the millisecond-protomagnetar stage, the second one is the BH-formation collapse phase and the last one is the Collapsar scenario with a Blandford-Znajek engine. 
Additionally, we analyze and explain the afterglow phase evoking the standard fireball model. 
Our model can successfully describe the
timescales, fluxes and spectral indices observed for GRB 110709B.}
\FullConference{Swift: 10 Years of Discovery,\\
		2-5 December 2014\\
		La Sapienza University, Rome, Italy }

\begin{document}

\section{Introduction}
The Collapsar model \cite{1993ApJ...405..273W,1999ApJ...524..262M} for long gamma-ray bursts (GRBs), requires the collapse of a stellar core with a specific angular momentum of $\sim$ $10^{16.5}$ cm$^2$ s$^{-1}$ \cite{2012ApJ...752...32W}. 
Under this condition the stellar core collapses to a black hole (BH) surrounded by an accretion disk. 
This implies pre-collapse spin periods of $P_{spin} \leq$ 0.5 day (and thus, orbital periods in tidally-synchronized binaries) which restricts the stellar evolution prior to collapse \cite{2002ApJ...575..996L,2007ApJ...671L..41B,2008ApJ...685.1063B,2008ApJ...689L...9M,2011MNRAS.413..183M,2014ApJ...781....3M}. 
On the other hand, Duncan \& Thompson \cite{1992ApJ...392L...9D} and later works \cite{2011MNRAS.413.2031M} have proposed utilizing the spin down of ms protomagnetars as the central engines for long GRBs. 
These engines have much less available rotational and binding energy when compared to collapsars (10$^{52}$ erg vs 10$^{54}$ erg, respectively) as the mass and spin of the compact object are both smaller; thus, they must be much more efficient in their energy conversion. 
Now, it is likely that magnetars may be produced by less massive stars than BHs, then they may be much more common.
% as the progenitor stars could be considerably less massive and, hence, much more abundant. 
%It is likely that, for these engines to work, . 
Nonetheless, it is still necessary for the progenitor of ms magnetars to rotate extremely rapidly prior to the collapse, both, to explain its ms rotation and, perhaps even, to amplify the magnetic field (e.g., \cite{2014ApJ...781....3M}). 
In this work we study the possibility, previously suggested in Zhang et al. \cite{2012ApJ...748..132Z}, that GRB 110709B was the result of a combination of factors that allowed both mechanisms, ms magnetar and collapsar, to play a part in this transient event.
\section{Model}
To explain the three peaks (or episodes) of GRB 110709B as the collapse of a single star we show the model as a function of time scales, magnetic fields, masses, etc.
\paragraph{First Episode:} We consider a collapsing $3.5 \msun$ core ($R_{Fe}\simeq 10^9$ cm). The free-fall timescale of such an object is of the order of 
\be
\tau_{ff} = \frac{\pi}{2} \sqrt{\frac{R_{Fe}^3}{GM_{Fe}}} \simeq \frac{3 \pi}{4} \;
\left(\frac{R}{10^9 {\rm cm}}\right)^{3/2}\left(\frac{M}{3.5\msun}\right)^{-1/2} {\rm s}. \label{eq:Kepler}
\ee
the viscous timescale, $\tau_v$, is
\be
\tau_v \simeq \left(\frac{r}{h}\right)^2\frac{4\tau_{ff}}{\alpha} \simeq 900 {\rm s},
\ee
and the rotational kinetic energy of a protoneutron star (PNS; or a neutron star, NS) is then of the order of
\be
E_k = \frac{1}{2}I \Omega^2 = \left(\frac{1}{2}\right)\left(\frac{2}{5}\right) \frac{G M^2}{R} %\nonumber\\
\simeq  133 \left(\frac{k}{2/5}\right) \left(\frac{M}{\msun}\right)^2 \left(\frac{10^6 {\rm cm}}{R}\right) {\rm B},
\ee
where a Bethe, $1$ B $= 10^{51}$ erg.
This energy be tapped through a torque exerted by the magnetic dipole. The power can be estimated from:
\be
\dot{E}_k \simeq \frac{2}{3}\frac{B^2 R^6 \Omega^4}{c^3} \simeq 2.2 \left(\frac{B}{3 \times 10^{14} {\rm G}}\right)^2\left(\frac{\Omega}{10^{3} {\rm s}^{-1}}\right)^4 {\rm B\; s}^{-1},
\ee
where $R$ is the radius of the magnetic dipole and $\Omega$ is the angular velocity (see Usov \cite{1992Natur.357..472U} for more details).  
After a few tens of seconds, the ms-magnetar engine slows down and the jets are shut down. 
Matter held out by propeller effect in an accretion disk will start streaming down, burying the magnetic field and increasing the mass of the NS.
\paragraph{Second Episode:}
By assuming that around ($k \simeq$) 10\% of the rest mass of the NS is released as binding energy during the conversion into a BH, the total energy available for the 2nd episode can be estimated from
\be
E_T \simeq k M_{NS}c^2 =  630 \left(\frac{k}{0.1}\right)\left(\frac{M_{NS}}{3.5\msun}\right) {\rm B}.
\ee
Under {\it normal} SN conditions, over $99 \%$ of the energy released leaves the star as neutrinos, without further interaction. For SN 1987A \cite{2014MNRAS.442..239F} the kinetic energy was around $E_{kin} \sim 1$ B and the total energy released (in neutrinos) was $E_T \sim 300$ B.  Assuming somewhat larger efficiency (larger density and temperature, thus, larger neutrino cross section), and considering we have more mass in the compact object we estimate the kinetic energy to be a few times larger than in SN 1987A, i.e., $E_{kin}  \sim 5$ B, which coincides with the observations for GRB 110709B. Given the available energy and the fact that the material that forms the collapsing magnetar is denser than that of the accretion disk, a quasi-thermalized, sharp, hard-X-ray signal would be expected. 

\paragraph{Third Episode:}
After the BH forms, it accretes material and angular momentum. 
As the rest of the stellar core collapses, differential rotation and convection regenerate a strong magnetic field. The Kerr BH interacts with the accretion disk via the Blandford-Znajek mechanism (BZ; \cite{1977MNRAS.179..433B}). The energy can be estimated from
\be
E_{BZ} = 1,800 \; \epsilon_\Omega f(a_\star) \left(\frac{M}{\msun}\right) {\rm B},
\ee
where $a_\star \equiv Jc/(GM^2) = a/M$, 
\be
f(a_\star) = 1 - \sqrt{\frac{1}{2}\left(1+\sqrt{1-a_\star^2}\right)},
\ee
and $\epsilon_\Omega = 0.5$, the maximum efficiency for energy extraction. The power of the BZ central engine is given by
\be
P_{BZ} \simeq 0.17 \;a_\star^2 \left(\frac{B}{10^{15}{\rm G}}\right)^2\left(\frac{M}{\msun}\right)^2 {\rm B \; s}^{-1},
\ee
(see \cite{2000PhR...325...83L,2011ApJ...727...29M} for more detail).

\section{Afterglow (AG)}
In the framework of the external shocks of the fireball model the deceleration radius and timescale are
\begin{equation}\label{eq:rdec}
R_{d}=\biggl(\frac{3}{4\pi\,m_p}\biggr)^{1/3}\,\Gamma^{-2/3}\,\eta^{-1/3}\,E^{1/3}\, ,
\end{equation}
and
\begin{equation}\label{eq:Tdec}
  t_{d}=\biggl(\frac{3}{32\pi\,m_p}\biggr)^{1/3}\,(1+z)\,\Gamma^{-8/3}\,\eta^{-1/3}\,E^{1/3}\,,
\end{equation}
respectively, where $E$ is the energy, $\Gamma$ is the bulk Lorentz factor, $\eta$ is the interstellar medium (ISM) density and $m_p$ is the proton mass. The AG emission is more likely to be of synchrotron origin, thus, the spectrum has a break in the fast cooling regime resulting in a spectral index of 1.5.

\begin{figure}
\hspace{-0.6cm}%\begin{center}
%\resizebox{15.6cm}\{!}{
\includegraphics[scale=0.65,angle=0]{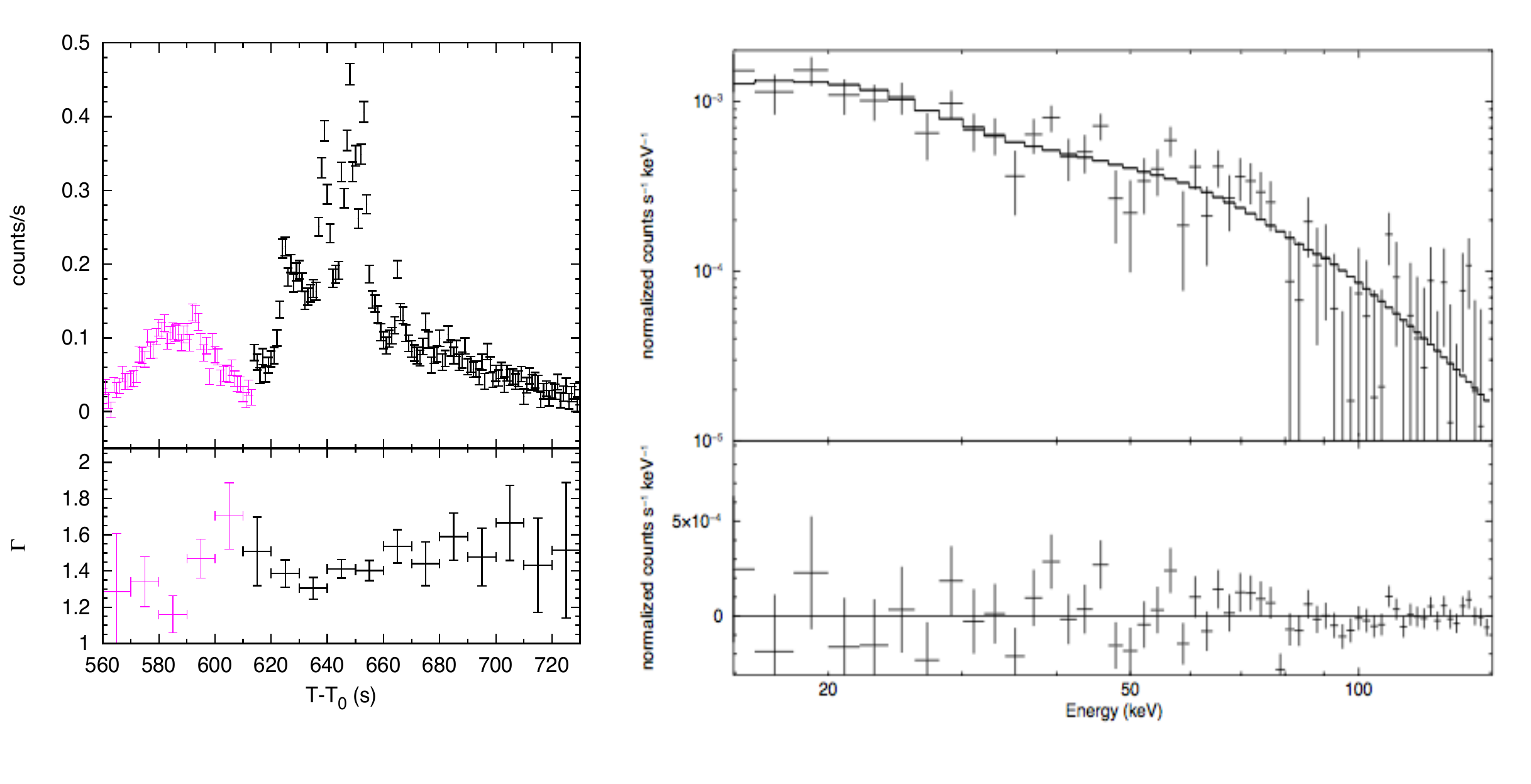}%}
\caption{{\it Left panel:}BAT count rates (upper panel) and photon index evolution (lower panel) during episodes 2 (in magenta) and 3 (in black) of GRB 110709B. The spectral model is a simple power law. The time is relative to the trigger time of episode 1, of 331940612 s (in MET seconds).
{\it Right panel:} Fit of the Swift/BAT spectrum with a PL model between 835 s and 865 s after the first trigger time. The lower panel shows the residuals.
}
\label{fig:LC}
%\end{center}
\end{figure}

%\begin{figure}
%\begin{center}
%%\resizebox{15.6cm}\{!}{
%\includegraphics[scale=1,angle=0]{LightCurve.pdf}%}
%\caption{BAT count rates (upper panel) and photon index evolution (lower panel) during episodes 2 (in magenta) and 3 (in black) of GRB 110709B. The spectral model is a simple power law. The time is relative to the trigger time of episode 1, of 331940612 s (in MET seconds).}
%\label{fig:LC}
%\end{center}
%\end{figure}

\section{Observations and Data Analysis}

%\begin{figure}
%\resizebox{8.8cm}{!}{\includegraphics[angle=0]{spectrum835-865.pdf}}
%\caption{Fit of the Swift/BAT spectrum between 835 s and 865 s after the trigger time of episode 1 with a PL model. The lower panel shows the residuals.}
%\label{fig:AG}
%\end{figure}

GRB 110709B triggered the Burst Alert Telescope (BAT, \cite{2005SSRv..120..143B}) onboard the Swift satellite at 21:32:39 UT on July 9, 2011 \cite{2011GCN..12144...1C}.  
This episode extended up to 55 s after the trigger \cite{2012ApJ...748..132Z}. Hereafter all events will be measured with reference to the first trigger time.  Surprisingly, there was a second BAT trigger at 21:43:25 UT on  July 9, 2011 after $\sim$ 11 minutes. 
The emission period extended up to 865 s \cite{2011GCN..12144...1C}; this period shows a first bump (episode 2), beginning at $\sim$ 550 s  and lasting about 60 s (ending at 610 s), followed by a second multi-peaked bump (episode 3, from 610 s to 750 s; see Fig. 1) of longer duration.  

Zhang et al. \cite{2012ApJ...748..132Z} performed a time-resolved spectral analysis of this second emission (episodes 2 and 3), dividing its time period in time slices of 50 s or more; they showed that the spectra can be fitted with cutoff power laws and that there was a strong hard-to-soft spectral evolution. 
However, their choice of the time intervals is not suitable to look at the evolution of the spectral parameters during episode 2, as the corresponding time slice covers its whole duration. 
Therefore, we performed a more detailed analysis of the spectra of episodes 2 and 3, by considering sub-intervals of 10 s each. 
We processed the \emph{Swift}/BAT data using the standard FTOOLS package (Heasoft, version 6.15). 
We analyzed the spectra using two different spectral models: BB+PL and PL. We found that all the spectra are well modeled with PL, while the BB+PL can be discarded, as it is not well constrained. 
The lower left panel in Fig. 1 shows the time evolution of the photon index of the PL model: it can be seen that a discontinuity in the hard-to-soft evolution comes out at the beginning of episode 3. 
This could suggest a different emission mechanism for episodes 2 and 3.  
We obtain that the BAT band (15-150) keV fluences of the first (from $-28$ to $55$ s), second ($550$ to $610$ s) and third ($610$ to $750$ s) episodes are $9.54^{+0.11}_{-0.16} \times 10^{-6}$ erg cm$^{-2}$, $2.31^{+0.06}_{-0.05} \times 10^{-6}$ erg cm$^{-2}$ and $8.81^{+0.09}_{-0.12} \times 10^{-6}$ erg cm$^{-2}$, respectively.

We also analyze the spectrum between 835 s and 865 s corresponding to the last peak observed in the BAT light curve \cite{2012ApJ...748..132Z}, tens of seconds after the end of episode 3.  We find that the BAT spectrum (see right panel in Fig.~\ref{fig:LC}) is well modeled with a PL with spectral photon index $\alpha_3 = 1.59^{+0.15}_{-0.14}$.   By considering the value of spectral photon index, the deceleration radius (eq. \ref{eq:rdec} ) and time scale (eq. \ref{eq:Tdec}),  we interpret that this peak is related to the AG of episode 3.
\section{Discussion and Conclusions}
Considering the values of the first episode: Energy E = 20 B, redshift z = 0.75 ($\sim$3.1 Gpc; see \cite{2013A&A...551A.133P}), ISM density $\eta$ = 3 cm$^{-3}$ and Lorentz factor of $\Gamma$ = 160 and $\Gamma$ = 50 we obtain deceleration radii of $R_d \sim 3.5 \times10^{16}$ cm and $R_d \sim 7.5\times 10^{16}$ cm which correspond to time scales of $t_d \sim$ 40 s and $t_d \sim$ 870 s, respectively.\\
Taking into consideration the values for episode 3, energy E = 18.5 B, a mean ISM density $\eta$= 10$^{-3}$ cm$^{-3}$ and a bulk Lorentz factor of $\Gamma$= 225, we obtain a value of deceleration time $t_d \sim$ 225s which is in agreement with the observations.\\
The first jet propagates out of the star and into the ISM clearing a path, thus lowering $\eta$ from 3 cm$^{-3}$ to 10$^{-3}$ cm$^{-3}$ . The first AG took place at 36 s ($\alpha_1$ = 1.55 $\pm$ 0.05 ; \cite{2012ApJ...748..132Z}), and the AG for 3rd episode occurred at 850 s ($\alpha_3$  = 1.59 $\pm$ 0.15)  \cite{2014arXiv1411.7377M}.
Zhang et al. (2012) \cite{2012ApJ...748..132Z} reported a component modeled with a power law of spectral index $\alpha_1 = 1.55\pm0.05$ at the very end of the prompt emission, between $\sim36$ and $\sim45$ s which strongly suggests an early AG probably powered by a magnetized outflow \cite{2012ApJ...751...33F, 2012ApJ...755..127S, Fraija}.

\end{document}